\documentclass[12pt,a4paper]{article}
\usepackage{epsfig}

\begin{document}

\begin{titlepage}
\begin{flushright}
     UPRF 97-01  \\
     January 1997 \\
\end{flushright}
\par \vskip 10mm
\begin{center}
{\Large \bf
Spin Networks and Recoupling in Loop 
Quantum Gravity\footnote{To appear in the Proceedings of the 2nd
Conference on Constrained Dynamics and Quantum Gravity, 
Santa Margherita, Italy, 17-21 September 1996.}$^,$\footnote{
This work has been partially supported by the INFN 
grant ``Iniziativa specifica FI-41'' (Italy).}
}
\end{center}
\par \vskip 2mm
\begin{center}
Roberto De Pietri
\vskip 5 mm
Dipartimento di Fisica, \\
Universit\`a degli studi di Parma \\
and \\
INFN, Gruppo Collegato di Parma, \\
ITALY.
\end{center}
\par \vskip 2mm\begin{center} {\large \bf Abstract} \end{center}
\begin{quote}
I discuss the role played by the spin-network basis and recoupling
theory (in its graphical tangle-theoretic formulation) and their
use for performing explicit calculations in loop quantum gravity.
In particular, I show that recoupling theory allows the
derivation of explicit expressions for the eingenvalues of the
quantum volume operator. An important side result of these
computations is the determination of a scalar product with respect
to which area and volume operators are symmetric, and the spin
network states are orthonormal.
\end{quote}
\end{titlepage}


\section{Introduction}

A promising attempt towards a quantum theory of
gravitation is provided by the loop quantization \cite{LOOP}.  It
amounts to the direct canonical quantization of the (Poisson bracket)
holonomy algebra generated by:
\begin{eqnarray}
  {\cal T}[\alpha]          &=& - {\rm Tr} [U_\alpha] ,
\label{t0}  \\
  {\cal T}^a[\alpha](s)     
      &=& - {\rm Tr} [U_\alpha(s,s) \tilde{E}^a(s)] ,
\end{eqnarray}
where $U_\alpha$ is the parallel propagator of the Ashtekar-Sen
connection $A_a$ along a loop $\alpha$. The crucial point is
that to each loop $\alpha$ is uniquely associated a ${\cal T}$
observable. This suggests to use as carrier space of the
representation a suitable subspace ${\cal V}$ of the free loop
algebra defined as the finite linear combinations
$\Phi$ of finite products of loops, as in:
\begin{equation}
\Phi = c_0 + \sum_i c_i [\alpha_i] + \sum_{jk} c_{jk}
                 [\alpha_j][\alpha_k]~ + \ldots ~.
\label{fla}
\end{equation}
In this talk I shall summarize the results obtained in \cite{DePietri96};
in particular, I shall show that the ${\cal V}$-space admits a
tangle-theoretic description in terms of the Temperley-Lieb recoupling
theory \cite{Kauffman94}, and that a convenient basis in this space is
given by the spin-network states.

It is important to note that an element of the ${\cal T}$ holonomy algebra
\[
{\cal T}[\Phi] 
    \! = \! c_0   
    \! + \!\sum_i c_i {\cal T}[\alpha_i]~
    \! + \!\sum_{ij} c_{ij} {\cal T}[\alpha_i]\,{\cal T}[\alpha_j]~
    \! + \ldots
\]
is associated  to each element of the free loop algebra, 
and that the connection representation is defined by introducing an
Hilbert space structure on a suitable extensions
$\overline{{\cal A}/{\cal G}}$ of this ${\cal T}$ algebra
\cite{CONNECTION}. The relation between the two representations
is given by the loop transform \cite{FunctINT}.  For a more detailed
analysis of this point see \cite{DePietri96a}.

\section{The Carrier Space of the Loop Representation and 
         the Spin-Networks Basis}

The loop representation of quantum gravity is assumed to be the
linear representation of the Poisson algebra of the ${\cal T}$
variables over ${\cal V}$ defined by: $\langle \Phi | ~{\cal
T}[\beta] = \langle \Phi \cdot [\beta] |$.  It may happen that
two elements $\Phi$ and $\Psi$ of the free loop algebra
correspond to the same function on the holonomy algebra ${\cal T}[\Phi]
={\cal T}[\Psi]$ (i.e., they give the same value for any values of
the connection).  As a consequence, the carrier space  ${\cal V}$ of the
representation must be defined as the space of the
equivalence classes of the free loop algebra under the equivalence
defined by all the relations (Mandelstam relations)
\begin{equation}
  \Phi\sim\Psi ~~~\mbox{if}~~~ {\cal T}[\Phi] ={\cal T}[\Psi],
\label{eq:Mandelstam}
\end{equation}
namely by the equality of the corresponding holonomies.
The principal consequences of the Mandelstam relations are that
the ${\cal V}$-space does not depend on the orientation and 
parameterization of the loops.
Moreover, the following identities hold true: 
{\bf [Retracing]}: if $\gamma$ is a {\it segment} starting in
a point of $\alpha$ then 
$[\alpha\circ\gamma\circ\gamma^{-1}]\sim [\alpha]$; 
{\bf [Binor identity]}: 
$[\alpha]\cdot[\beta]\sim-[\alpha\#_s\beta] - [\alpha \#_s \beta^{-1}]$.
 
According to \cite{DePietri96}, a natural way to represent an element $\Phi$
of ${\cal V}$ is the following: first, one introduces the extended planar
graph $\Gamma_{ex}$ of $\Phi$ as the two-dimensional surface obtained by
``thickening out'' a planar representation of the image $\gamma$ of all
the loop in $\Phi$; then, one draws the loops of $\Phi$ over
$\Gamma_{ex}$.  In this way, to each element of the ${\cal V}$-space
corresponds an element in the algebra of all possible tangles over
$\Gamma_{ex}$. Moreover, in the tangle-theoretic planar representation of
the states ${\langle \Phi |}$, the {\it retracing} and {\it binor}
identities become the following tangle identities
\begin{eqnarray}
&&
\begin{array}{c}{\setlength{\unitlength}{1 pt}
\begin{picture}(10,14) 
   \put( 5,7){\oval(10,14)}
\end{picture}}\end{array}
= - 2 
\\
&&
\begin{array}{c}{\setlength{\unitlength}{1 pt}
\begin{picture}(10,15) 
\put( 0,0){\line( 2,3){10}}
\put(10,0){\line(-2,3){10}}
\end{picture}}\end{array}
= - \begin{array}{c}{\setlength{\unitlength}{1 pt}
\begin{picture}(10,15) 
\put( 0,0){\line(0,1){15}}
\put(10,0){\line(0,1){15}}
\end{picture}}\end{array}
 - \begin{array}{c}{\setlength{\unitlength}{1 pt}
\begin{picture}(10,15) 
\put(5, 0){\oval(10,10)[t]}
\put(5,15){\oval(10,10)[b]} 
\end{picture}}\end{array}
\label{gbi} 
~~.
\end{eqnarray}
These identities are the key axioms of the tangle-theoretic
recoupling theory \cite{Kauffman94}, if the value of the $A$
parameter set to $-1$.  This gives the possibility of using the whole
machinery of the tangle-theoretic recoupling theory
\cite{Kauffman94} for dealing with the graphical representation of 
the states.  In particular, a basis in this tangle-theoretic algebra is
provided by the spin-networks, where a {spin network} $S$ is the sum 
of tangles given as follow: consider a three-valent ``virtual'' 
graph $\Gamma^v$ over $\Gamma_{ex}$ and a ``{\it compatible
coloring}'' $\{p_e\}$ of the  edges of $\Gamma^v$; replace each
edge with the full-antisymmetric sum of $p_e$ parallel lines;
then, in each three-valent vertices connect all the incoming
lines in the unique possible planar way. 

\begin{figure}[t]
\begin{center}
\mbox{\epsfig{file=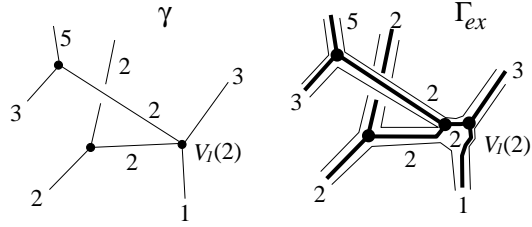,width=7cm}}
\end{center}
\caption{An example of spin-network state and of its 
extended planar representation}
\label{fig:SN}
\end{figure}

In this way, any element of the ${\cal V}$-space can be characterized in
terms of the tangle-theoretic {\bf spin network quantum state}
${\left\langle{{S}}\right|}=(\gamma,\Gamma_{ex},S)$, defined as the
element of $\cal V$ determined by the graph $\gamma$, its extended
planar graph $\Gamma_{ex}$, and the spin-network tangle $S$ over
$\Gamma_{ex}$.  These states are essentially characterized by the 
graph $\gamma$,
the number $p_e$ of loops in each edge $e$ of $\gamma$, and the
recoupling (in terms of three-valent vertices) of the edges
connected to each vertex of $\gamma$ (see figure \ref{fig:SN}).

Note that, representing a vertex of a spin network state $\langle S|$ by
means of the portion of its spin-network $S$ contained in the vertex
\begin{equation} \label{eq:vert}
\left\langle V^{(n)}[P_0,\ldots,P_{n-1}]({K}_{I_i}) \right|
   = 
\Bigg\langle{\!\!\! \begin{array}{c}
   \setlength{\unitlength}{1 pt}
   \begin{picture}(120,55)
  \put( 0, 0){$P_0$}\put(15, 0){\line(1,1){10}}
  \put( 0,20){$P_1$}\put(15,20){\line(1,0){10}}
  \put(10,40){$P_2$}\put(25,40){\line(1,-1){10}}
  \put(25,10){\line(0,1){10}} \put(27,10){${\scriptstyle  i_1}$}
  \put(25,20){\line(1,1){10}} \put(32,18){${\scriptstyle  i_2}$}
  \put(35,30){\line(1,0){10}} \put(40,22){${\scriptstyle  i_3}$}
  \put(25,20){\circle*{3}}\put(35,30){\circle*{3}}
  \put(50,40){$\cdots$}        \put(50,30){$\ldots$} 
  \put(70,30){\line(1, 0){10}} \put(58,20){$ $} 
  \put(80,30){\line(1,-1){10}} \put(65,20){${\scriptstyle  i_{n-2}}$}
  \put(90,20){\line(0,-1){10}} \put(72, 8){${\scriptstyle  i_{n-1}}$}
  \put(90,20){\circle*{3}}\put(80,30){\circle*{3}}
  \put(95,40){$P_{n-3}$}\put(80,30){\line(1,1){10}}
  \put(105,20){$P_{n-2}$}\put(90,20){\line(1,0){10}}
  \put(105, 0){$P_{n-1}$}\put(90,10){\line(1,-1){10}} 
\end{picture}\end{array} 
~~}\Bigg| ~~,
\end{equation}
and considering two vertices, $V_n[a_1,\ldots,a_n]$ and
$V_n^\prime[b_1,\ldots,b_n]$, they can connected only if all the
corresponding incoming lines have the same colors.  In this case, it is
possible to attach the $a_i$ lines and the corresponding $b_i$ lines
to each other. Then, it is obtained a tanle over $\Gamma_{ex}$
constituted only of loops contractible to a point. By recoupling, 
a tangle in which all the loops are contractible to a point
reduce to a number. The operation of computing this number 
is named chromatic evaluation.  This evaluation,
denoted by
${\langle}V_n(a_1,\ldots,a_n)|V_n^\prime(b_1,\ldots,b_n)\rangle$, is
a scalar product in the space of the possible vertices.  In
particular, the chromatic evaluation of a $2$-vertex (i.e. of a
line) and that of a $3$-vertex will be denoted as $\Delta$ function
and $\theta$ function, respectively:
\begin{eqnarray}
  \Delta_n &=& \langle V_2[n],V_2[n] \rangle \! 
  =\!\begin{array}{c}\setlength{\unitlength}{.5 pt}
     \begin{picture}(35,25)
        \put(15, 0){\line(0,1){10}}\put(20, 0){\line(0,1){10}}
        \put(15, 0){\line(1,0){ 5}}\put(15,10){\line(1,0){ 5}}
        \put(15,25){\line(1,0){5}}  \put(15,15){$\scriptstyle n$}
        \put(15,15){\oval(30,20)[l]}\put(20,15){\oval(30,20)[r]} 
     \end{picture}\end{array} \!
\\
\theta(a,b,c) 
  &=&  \langle V_3[a,b,c],V_3[a,b,c] \rangle 
   = \begin{array}{c}\setlength{\unitlength}{.5 pt}
     \begin{picture}(40,40)
        \put(18,32){$\scriptstyle a$}
        \put( 0,15){\line(1,0){40}} \put(18,17){$\scriptstyle b$}
        \put(20,15){\oval(40,30)}   \put(18, 2){$\scriptstyle c$}
        \put( 0,15){\circle*{3}}    \put(40,15){\circle*{3}}
     \end{picture}\end{array}
\! .
\end{eqnarray} 
For the explicit values of this evaluation and all the details 
see \cite{DePietri96} and \cite{Kauffman94}.
In this tangle-theoretic interpretation of the loop representation,
one still has the analogous of the Wigner-Eckart theorem: 
the recoupling theorem of \cite{Kauffman94} (pg.\ 60)
states, as a tangle relation, that
\begin{equation}
\begin{array}{c}\setlength{\unitlength}{1 pt}
\begin{picture}(50,40)
          \put( 0,0){$a$}\put( 0,30){$b$}
          \put(45,0){$d$}\put(45,30){$c$}
          \put(10,10){\line(1,1){10}}\put(10,30){\line(1,-1){10}}
          \put(30,20){\line(1,1){10}}\put(30,20){\line(1,-1){10}}
          \put(20,20){\line(1,0){10}}\put(22,25){$j$}
          \put(20,20){\circle*{3}}\put(30,20){\circle*{3}}
\end{picture}\end{array}
 \!\!   = \sum_i \! 
       \left\{\begin{array}{ccc}
                      a  & b & i \\
                      c  & d & j  
              \end{array}\right\} \!\!
\begin{array}{c}\setlength{\unitlength}{1 pt}
\begin{picture}(40,40)
      \put( 0,0){$a$}\put( 0,40){$b$}
      \put(35,0){$d$}\put(35,40){$c$}
      \put(10,10){\line(1,1){10}}\put(10,40){\line(1,-1){10}}
      \put(20,30){\line(1,1){10}}\put(20,20){\line(1,-1){10}}
      \put(20,20){\line(0,1){10}}\put(22,22){$i$}
      \put(20,20){\circle*{3}}\put(20,30){\circle*{3}}
\end{picture}\end{array}
\!\! ,
\end{equation}
where the quantities $\left\{  {}^{a~b~i}_{c~d~j} \right\}$ are $su(2)$
six-j symbols (normalized as in \cite{Kauffman94}).

One of the main advantages of the spin-network basis is that
the action of the ${\cal T}^a[\alpha](s)$ operators is 
particularly simple. In fact, one has:
\begin{equation}
\mbox{\epsfig{file=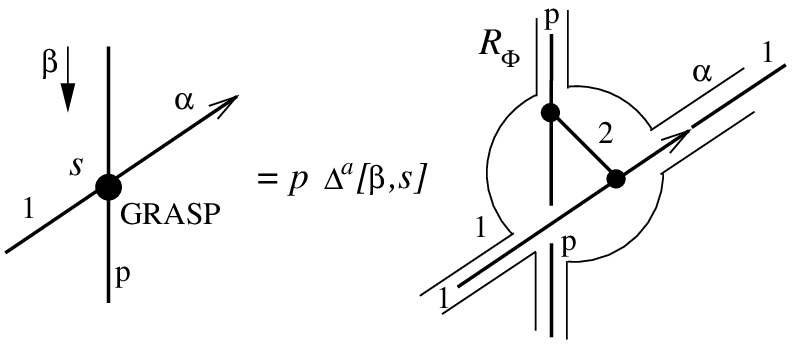,width=7cm}}
\!\!\!
\label{eq:THEGRASP}
\end{equation}
where $\Delta^a[\beta,s]=\int_\beta d\lambda \dot{\beta}^a(\lambda)
\delta^3\![\beta(\lambda),s]$.  Indeed the action of the 
operators corresponds to adding a new vertex of the kind shown in
eq. (\ref{eq:THEGRASP}) at the intersection point $s$ between $\alpha$ and
$\beta$. This gives the possibility of expressing 
the action of any operator on the spin network basis as a product of 
three parts: an analytic  regularization factor, a graphical part (due the
a possible change of $\gamma$),  and a vertex part (see \cite{Borissov}). 
The vertex part, connected with the reduction to elementary 
vertices, can be computed using the recoupling theory and the 
associated chromatic evaluations. 

\section{The Volume Operator}

As an example of application of the recoupling theory, I show how is
possible to obtain explicit formulas for the eingenvalues of the volume
operator. In \cite{DePietri96}, it has been shown that the action of the
volume operator is given by a regularization factor and a vertex
part. The following final expression for the action of the volume
operator on a given
spin-network has been obtained
\begin{eqnarray}
 \hat V[{\cal V}] &=& 
 \sum_{i\in\{{S\cap{\cal V}}\}} \hat V_i, 
\nonumber \\
\hat V_i &=&  {l_0^3}
\sqrt{  \sum_{\scriptstyle  r\neq s\neq t=0}^{n_i-1}
  \left| \frac{{\rm i}}{16~3!}~ \hat W_{[rts]}^{(n_i)} ~\right|} 
~~.
\label{eq:VolOpSN2}
\end{eqnarray}
The first sum is over all the vertices and the second sum is
over the triples of edges adjacent to each vertex.  
In \cite{DePietri96}, it has 
also been shown that the vertex operators $\hat W_{[rts]}^{(n_i)}$ are
represented by diagonalizable matrices with real eigenvalues. 
In fact, a normalization of the vertex exists such that, in this basis,
the vertex operator $\hat W_{[rts]}^{(n_i)}$ are represented by 
the real and antisymmetric matrices 
\[
\left\langle \! V^{(\! n_i\! )}({K}_{I_i}) \right|_N  
   \hat W_{[rts]}^{(n_i)}
 = \!\! \sum_{\bar{K}_{I_i}} 
   \hat W_{[rts]}^{(n_i)}{}_{K_{I_i}}^{\bar{K}_{I_i}}
   \left\langle \! V^{(\! n_i \!)}(\bar{K}_{I_i}) \right|_N 
\!\!\! .
\]
In the case of 4-valent vertices, this normalization is explicitly given by:
\begin{equation}
\left\langle V^{(4)}(i) \right|_N =  \sqrt{\frac{
     \begin{array}{c}\setlength{\unitlength}{.5 pt}
     \begin{picture}(35,25)
        \put(15, 0){\line(0,1){10}}\put(20, 0){\line(0,1){10}}
        \put(15, 0){\line(1,0){ 5}}\put(15,10){\line(1,0){ 5}}
        \put(15,25){\line(1,0){5}}  \put(15,15){$\scriptstyle i$}
        \put(15,15){\oval(30,20)[l]}\put(20,15){\oval(30,20)[r]}
     \end{picture}\end{array}
   }{\begin{array}{c}\setlength{\unitlength}{.5 pt}\begin{picture}(40,40)
        \put(18,32){$\scriptstyle a$}
        \put(18,17){$\scriptstyle b$} 
        \put(18, 2){$\scriptstyle i$} 
        \put(20,15){\oval(40,30)} \put( 0,15){\line(1,0){40}} 
        \put( 0,15){\circle*{3}}  \put(40,15){\circle*{3}}
     \end{picture}\end{array}
     \begin{array}{c}\setlength{\unitlength}{.5 pt}\begin{picture}(40,40)
        \put(18,32){$\scriptstyle c$}
        \put(18,17){$\scriptstyle d$} 
        \put(18, 2){$\scriptstyle i$} 
        \put(20,15){\oval(40,30)} \put( 0,15){\line(1,0){40}} 
        \put( 0,15){\circle*{3}}  \put(40,15){\circle*{3}}
     \end{picture}\end{array}
   }} 
\left\langle \!\!\!
\begin{array}{c}\setlength{\unitlength}{1 pt}
\begin{picture}(40,40)
          \put( 0,0){$a$}\put( 0,30){$b$}
          \put(35,0){$d$}\put(35,30){$c$}
          \put( 5,10){\line(1,1){10}}\put( 5,30){\line(1,-1){10}}
          \put(25,20){\line(1,1){10}}\put(25,20){\line(1,-1){10}}
          \put(15,20){\line(1,0){10}}\put(17,25){$i$}
          \put(15,20){\circle*{3}}   \put(25,20){\circle*{3}}
\end{picture}\end{array}
\!\! \right|
~,
\label{eq:newbasis}
\end{equation}
and the matrix elements of the vertex operator are given by the
chromatic evaluation:
\begin{equation} \label{eq:14}
\tilde{W}^{(4)}_{[012]}(a,b,c,d){}_i^k = 
a\cdot b \cdot c\cdot \!
\sqrt{\frac{
     \begin{array}{c}\setlength{\unitlength}{.5 pt}
     \begin{picture}(35,25)
        \put(15, 0){\line(0,1){10}}\put(20, 0){\line(0,1){10}}
        \put(15, 0){\line(1,0){ 5}}\put(15,10){\line(1,0){ 5}}
        \put(15,25){\line(1,0){5}}  \put(15,15){$\scriptstyle i$}
        \put(15,15){\oval(30,20)[l]}\put(20,15){\oval(30,20)[r]}
     \end{picture}\end{array}
     \begin{array}{c}\setlength{\unitlength}{.5 pt}
     \begin{picture}(35,25)
        \put(15, 0){\line(0,1){10}}\put(20, 0){\line(0,1){10}}
        \put(15, 0){\line(1,0){ 5}}\put(15,10){\line(1,0){ 5}}
        \put(15,25){\line(1,0){5}}  \put(15,15){$\scriptstyle k$}
        \put(15,15){\oval(30,20)[l]}\put(20,15){\oval(30,20)[r]}
     \end{picture}\end{array}
   }{\begin{array}{c}\setlength{\unitlength}{.5 pt}\begin{picture}(40,40)
        \put(18,32){$\scriptstyle a$}
        \put(18,17){$\scriptstyle b$} 
        \put(18, 2){$\scriptstyle i$} 
        \put(20,15){\oval(40,30)} \put( 0,15){\line(1,0){40}} 
        \put( 0,15){\circle*{3}}  \put(40,15){\circle*{3}}
     \end{picture}\end{array}
     \begin{array}{c}\setlength{\unitlength}{.5 pt}\begin{picture}(40,40)
        \put(18,32){$\scriptstyle a$}
        \put(18,17){$\scriptstyle b$} 
        \put(18, 2){$\scriptstyle k$} 
        \put(20,15){\oval(40,30)} \put( 0,15){\line(1,0){40}} 
        \put( 0,15){\circle*{3}}  \put(40,15){\circle*{3}}
     \end{picture}\end{array}
     \begin{array}{c}\setlength{\unitlength}{.5 pt}\begin{picture}(40,40)
        \put(18,32){$\scriptstyle c$}
        \put(18,17){$\scriptstyle d$} 
        \put(18, 2){$\scriptstyle i$} 
        \put(20,15){\oval(40,30)} \put( 0,15){\line(1,0){40}} 
        \put( 0,15){\circle*{3}}  \put(40,15){\circle*{3}}
     \end{picture}\end{array}
     \begin{array}{c}\setlength{\unitlength}{.5 pt}\begin{picture}(40,40)
        \put(18,32){$\scriptstyle c$}
        \put(18,17){$\scriptstyle d$} 
        \put(18, 2){$\scriptstyle k$} 
        \put(20,15){\oval(40,30)} \put( 0,15){\line(1,0){40}} 
        \put( 0,15){\circle*{3}}  \put(40,15){\circle*{3}}
     \end{picture}\end{array}
   }} 
\cdot \!\!
\begin{array}{c}
  \setlength{\unitlength}{1 pt}
  \begin{picture}(110,60)
  \put(15,40){\line(0,1){10}}  \put(17,38){$a$}
  \put(15,32){\oval(20,20)[tl]} 
  \put(15,42){\circle*{3}}    
  \put(40,30){\line(0,1){30}}  \put(42,38){$b$}
  \put(40,32){\oval(20,20)[tl]} 
  \put(40,42){\circle*{3}}     \put(40,30){\circle*{3}}
  \put(40,60){\circle*{3}}
  \put(65,30){\line(0,1){30}}  \put(67,38){$c$}
  \put(65,32){\oval(20,20)[tl]} 
  \put(65,42){\circle*{3}}     \put(65,30){\circle*{3}}
  \put(65,60){\circle*{3}}
  \put(90,40){\line(0,1){10}}  \put(92,38){$d$}
  \put( 5,32){\line(0,-1){12}} \put(15, 2){${\scriptstyle  2}$}
  \put(30,28){\line(0,-1){18}} \put(22,15){${\scriptstyle  2}$}
  \put(55,28){\line(0,-1){ 8}} \put(40, 2){${\scriptstyle  2}$}
  \put(30,20){\oval(50,20)[b]} \put(30,10){\circle*{3}}
  \put(40,40){\oval(50,20)[bl]}\put(65,40){\oval(50,20)[br]}
  \put(40,30){\line(1,0){20}}  \put(42,20){$i$} 
  \put(40,50){\oval(50,20)[tl]}\put(65,50){\oval(50,20)[tr]}
  \put(40,60){\line(1,0){20}}  \put(52,50){$k$}
\end{picture}\end{array}  \!\!\!\! .
\end{equation}
The matrices $\tilde{W}^{(4)}_{[012]}(a,b,c,d){}_i^k$
have elements different from zero only if $|i-k|=2$. 
Using the equations
\begin{equation}
\label{eq:TETred}
\begin{array}{c}
  \setlength{\unitlength}{1 pt}
  \begin{picture}(40,40)
  \put(10, 5){\line(1,0){10}}   \put(2, 5){$i$}
  \put(10,20){\line(1,0){10}}   \put(2,20){$a$}
  \put(10,35){\line(1,0){10}}   \put(2,35){$k$}
  \put(20, 5){\circle*{3}}      
  \put(20,20){\circle*{3}}      
  \put(20,35){\circle*{3}}      
  \put(20, 5){\line(0,1){15}}   \put(22,10){$\bar{c}$}
  \put(20,20){\line(0,1){15}}   \put(22,25){$c$}
  \put(20,20){\oval(20,30)[r]}  \put(32,25){$d$}
\end{picture}\end{array}  
= \frac{\begin{array}{c}\setlength{\unitlength}{.8 pt}
        \begin{picture}(55,40)
        \put( 0,15){\line(1,-1){15}} \put(0,22){${\scriptstyle k}$}
        \put( 0,15){\line(1, 1){15}} \put(0, 0){${\scriptstyle i}$}
        \put( 0,15){\circle*{3}}
        \put(30,15){\line(-1, 1){15}} \put(28,20){${\scriptstyle c}$}
        \put(30,15){\line(-1,-1){15}} \put(28, 2){${\scriptstyle \bar{c}}$}
        \put(30,15){\circle*{3}}
        \put( 0,15){\line(1,0){30}} \put(12,16){${\scriptstyle a}$}
        \put(15,30){\line(1,0){25}} \put(15,30){\circle*{3}}
        \put(15, 0){\line(1,0){25}} \put(15, 0){\circle*{3}}
        \put(40, 0){\line(0,1){30}} \put(42,12){${\scriptstyle d}$}
        \end{picture}\end{array}
    }{ \begin{array}{c}\setlength{\unitlength}{.5 pt}\begin{picture}(40,40)
            \put(18,32){$\scriptstyle k$}
            \put(18,17){$\scriptstyle a$} 
            \put(18, 2){$\scriptstyle i$} 
            \put(20,15){\oval(40,30)} \put( 0,15){\line(1,0){40}} 
            \put( 0,15){\circle*{3}}  \put(40,15){\circle*{3}}
           \end{picture}\end{array}
      } 
~\cdot~
\begin{array}{c}
  \setlength{\unitlength}{1 pt}
  \begin{picture}(20,40)
  \put(2, 5){$i$}
  \put(10,20){\line(1,0){15}}   \put(2,20){$a$}
  \put(2,35){$k$}
  \put(25,20){\circle*{3}}      
  \put(10,20){\oval(30,30)[r]} 
\end{picture}\end{array}  
\end{equation}
\begin{equation}
\label{eq:MOVE2}
r ~\cdot \!
\begin{array}{c}
  \setlength{\unitlength}{1 pt}
  \begin{picture}(30,40)
  \put(15,20){\line(0,-1){10}}
  \put(15,20){\line( 1,1){10}}
  \put(15,20){\line(-1,1){10}}
  \put(15,20){\circle*{3}}
  \put( 3,32){${\scriptstyle p}$}
  \put(23,32){${\scriptstyle q}$}
  \put(13, 2){${\scriptstyle r}$}
  \put(15,15){\line(-1,0){7}}
  \put(15,15){\circle*{3}}
  \put( 6,16){${\scriptstyle 2}$}
\end{picture}\end{array}  
= p~\cdot  \!
\begin{array}{c}
  \setlength{\unitlength}{1 pt}
  \begin{picture}(30,40)
  \put(15,20){\line(0,-1){10}}
  \put(15,20){\line( 1,1){10}}
  \put(15,20){\line(-1,1){10}}
  \put(15,20){\circle*{3}}
  \put( 7,32){${\scriptstyle p}$}
  \put(23,32){${\scriptstyle q}$}
  \put(13, 2){${\scriptstyle r}$}
  \put(10,25){\line(-1,0){7}}
  \put(10,25){\circle*{3}}
  \put( 1,26){${\scriptstyle 2}$}
\end{picture}\end{array}  
+ q~\cdot \!
\begin{array}{c}
  \setlength{\unitlength}{1 pt}
  \begin{picture}(30,40)
  \put(15,20){\line(0,-1){10}}
  \put(15,20){\line( 1,1){10}}
  \put(15,20){\line(-1,1){10}}
  \put(15,20){\circle*{3}}
  \put( 3,32){${\scriptstyle p}$}
  \put(23,32){${\scriptstyle q}$}
  \put(13, 2){${\scriptstyle r}$}
  \put(20,25){\line(-1,0){7}}
  \put(20,25){\circle*{3}}
  \put(11,26){${\scriptstyle 2}$}
\end{picture}\end{array}  
\! ~,
\end{equation}
it is possible to reduce eq.\ (\ref{eq:14}) to the following 
chromatic evaluation:
\begin{eqnarray} \label{eqWred}
\begin{array}{l}
\tilde{W}^{(4)}_{[012]}(a,b,c,d){}_i^k =
   \begin{array}{c}\setlength{\unitlength}{.5 pt}\begin{picture}(40,40)
            \put(18,32){$\scriptstyle k$}
            \put(18,17){$\scriptstyle 2$} 
            \put(18, 2){$\scriptstyle i$} 
            \put(20,15){\oval(40,30)} \put( 0,15){\line(1,0){40}} 
            \put( 0,15){\circle*{3}}  \put(40,15){\circle*{3}}
   \end{picture}\end{array}
   ~\cdot~\sqrt{\frac{
     \begin{array}{c}\setlength{\unitlength}{.5 pt}
     \begin{picture}(35,25)
        \put(15, 0){\line(0,1){10}}\put(20, 0){\line(0,1){10}}
        \put(15, 0){\line(1,0){ 5}}\put(15,10){\line(1,0){ 5}}
        \put(15,25){\line(1,0){5}}  \put(15,15){$\scriptstyle i$}
        \put(15,15){\oval(30,20)[l]}\put(20,15){\oval(30,20)[r]}
     \end{picture}\end{array}
     \begin{array}{c}\setlength{\unitlength}{.5 pt}
     \begin{picture}(35,25)
        \put(15, 0){\line(0,1){10}}\put(20, 0){\line(0,1){10}}
        \put(15, 0){\line(1,0){ 5}}\put(15,10){\line(1,0){ 5}}
        \put(15,25){\line(1,0){5}}  \put(15,15){$\scriptstyle k$}
        \put(15,15){\oval(30,20)[l]}\put(20,15){\oval(30,20)[r]}
     \end{picture}\end{array}
   }{\begin{array}{c}\setlength{\unitlength}{.5 pt}\begin{picture}(40,40)
        \put(18,32){$\scriptstyle a$}
        \put(18,17){$\scriptstyle b$} 
        \put(18, 2){$\scriptstyle i$} 
        \put(20,15){\oval(40,30)} \put( 0,15){\line(1,0){40}} 
        \put( 0,15){\circle*{3}}  \put(40,15){\circle*{3}}
     \end{picture}\end{array}
     \begin{array}{c}\setlength{\unitlength}{.5 pt}\begin{picture}(40,40)
        \put(18,32){$\scriptstyle a$}
        \put(18,17){$\scriptstyle b$} 
        \put(18, 2){$\scriptstyle k$} 
        \put(20,15){\oval(40,30)} \put( 0,15){\line(1,0){40}} 
        \put( 0,15){\circle*{3}}  \put(40,15){\circle*{3}}
     \end{picture}\end{array}
     \begin{array}{c}\setlength{\unitlength}{.5 pt}\begin{picture}(40,40)
        \put(18,32){$\scriptstyle c$}
        \put(18,17){$\scriptstyle d$} 
        \put(18, 2){$\scriptstyle i$} 
        \put(20,15){\oval(40,30)} \put( 0,15){\line(1,0){40}} 
        \put( 0,15){\circle*{3}}  \put(40,15){\circle*{3}}
     \end{picture}\end{array}
     \begin{array}{c}\setlength{\unitlength}{.5 pt}\begin{picture}(40,40)
        \put(18,32){$\scriptstyle c$}
        \put(18,17){$\scriptstyle d$} 
        \put(18, 2){$\scriptstyle k$} 
        \put(20,15){\oval(40,30)} \put( 0,15){\line(1,0){40}} 
        \put( 0,15){\circle*{3}}  \put(40,15){\circle*{3}}
     \end{picture}\end{array}
   }}  ~\cdot~ 
\\ ~~~ \cdot
\bigg[ b ~
\frac{\begin{array}{c}\setlength{\unitlength}{.8 pt}
        \begin{picture}(55,40)
        \put( 0,15){\line(1,-1){15}} \put(0,22){${\scriptstyle 2}$}
        \put( 0,15){\line(1, 1){15}} \put(0, 0){${\scriptstyle 2}$}
        \put( 0,15){\circle*{3}}
        \put(30,15){\line(-1, 1){15}} \put(28,20){${\scriptstyle b}$}
        \put(30,15){\line(-1,-1){15}} \put(28, 2){${\scriptstyle b}$}
        \put(30,15){\circle*{3}}
        \put( 0,15){\line(1,0){30}} \put(12,16){${\scriptstyle 2}$}
        \put(15,30){\line(1,0){25}} \put(15,30){\circle*{3}}
        \put(15, 0){\line(1,0){25}} \put(15, 0){\circle*{3}}
        \put(40, 0){\line(0,1){30}} \put(42,12){${\scriptstyle b}$}
        \end{picture}\end{array}
    }{ \begin{array}{c}\setlength{\unitlength}{.5 pt}\begin{picture}(40,40)
            \put(18,32){$\scriptstyle b$}
            \put(18,17){$\scriptstyle b$} 
            \put(18, 2){$\scriptstyle 2$} 
            \put(20,15){\oval(40,30)} \put( 0,15){\line(1,0){40}} 
            \put( 0,15){\circle*{3}}  \put(40,15){\circle*{3}}
           \end{picture}\end{array}
      } 
 - k ~
\frac{\begin{array}{c}\setlength{\unitlength}{.8 pt}
        \begin{picture}(55,40)
        \put( 0,15){\line(1,-1){15}} \put(0,22){${\scriptstyle 2}$}
        \put( 0,15){\line(1, 1){15}} \put(0, 0){${\scriptstyle 2}$}
        \put( 0,15){\circle*{3}}
        \put(30,15){\line(-1, 1){15}} \put(28,20){${\scriptstyle k}$}
        \put(30,15){\line(-1,-1){15}} \put(28, 2){${\scriptstyle i}$}
        \put(30,15){\circle*{3}}
        \put( 0,15){\line(1,0){30}} \put(12,16){${\scriptstyle 2}$}
        \put(15,30){\line(1,0){25}} \put(15,30){\circle*{3}}
        \put(15, 0){\line(1,0){25}} \put(15, 0){\circle*{3}}
        \put(40, 0){\line(0,1){30}} \put(42,12){${\scriptstyle k}$}
        \end{picture}\end{array}
    }{ \begin{array}{c}\setlength{\unitlength}{.5 pt}\begin{picture}(40,40)
            \put(18,32){$\scriptstyle k$}
            \put(18,17){$\scriptstyle 2$} 
            \put(18, 2){$\scriptstyle i$} 
            \put(20,15){\oval(40,30)} \put( 0,15){\line(1,0){40}} 
            \put( 0,15){\circle*{3}}  \put(40,15){\circle*{3}}
           \end{picture}\end{array}
      } 
\bigg] \cdot\bigg[ b ~
\frac{\begin{array}{c}\setlength{\unitlength}{.8 pt}
        \begin{picture}(55,40)
        \put( 0,15){\line(1,-1){15}} \put(0,22){${\scriptstyle k}$}
        \put( 0,15){\line(1, 1){15}} \put(0, 0){${\scriptstyle i}$}
        \put( 0,15){\circle*{3}}
        \put(30,15){\line(-1, 1){15}} \put(28,20){${\scriptstyle b}$}
        \put(30,15){\line(-1,-1){15}} \put(28, 2){${\scriptstyle b}$}
        \put(30,15){\circle*{3}}
        \put( 0,15){\line(1,0){30}} \put(12,16){${\scriptstyle 2}$}
        \put(15,30){\line(1,0){25}} \put(15,30){\circle*{3}}
        \put(15, 0){\line(1,0){25}} \put(15, 0){\circle*{3}}
        \put(40, 0){\line(0,1){30}} \put(42,12){${\scriptstyle a}$}
        \end{picture}\end{array}
    }{ \begin{array}{c}\setlength{\unitlength}{.5 pt}\begin{picture}(40,40)
            \put(18,32){$\scriptstyle k$}
            \put(18,17){$\scriptstyle 2$} 
            \put(18, 2){$\scriptstyle i$} 
            \put(20,15){\oval(40,30)} \put( 0,15){\line(1,0){40}} 
            \put( 0,15){\circle*{3}}  \put(40,15){\circle*{3}}
           \end{picture}\end{array}
      } 
\bigg] \!\cdot\! \bigg[ c ~
\frac{\begin{array}{c}\setlength{\unitlength}{.8 pt}
        \begin{picture}(55,40)
        \put( 0,15){\line(1,-1){15}} \put(0,22){${\scriptstyle k}$}
        \put( 0,15){\line(1, 1){15}} \put(0, 0){${\scriptstyle i}$}
        \put( 0,15){\circle*{3}}
        \put(30,15){\line(-1, 1){15}} \put(28,20){${\scriptstyle c}$}
        \put(30,15){\line(-1,-1){15}} \put(28, 2){${\scriptstyle c}$}
        \put(30,15){\circle*{3}}
        \put( 0,15){\line(1,0){30}} \put(12,16){${\scriptstyle 2}$}
        \put(15,30){\line(1,0){25}} \put(15,30){\circle*{3}}
        \put(15, 0){\line(1,0){25}} \put(15, 0){\circle*{3}}
        \put(40, 0){\line(0,1){30}} \put(42,12){${\scriptstyle d}$}
        \end{picture}\end{array}
    }{ \begin{array}{c}\setlength{\unitlength}{.5 pt}\begin{picture}(40,40)
            \put(18,32){$\scriptstyle k$}
            \put(18,17){$\scriptstyle 2$} 
            \put(18, 2){$\scriptstyle i$} 
            \put(20,15){\oval(40,30)} \put( 0,15){\line(1,0){40}} 
            \put( 0,15){\circle*{3}}  \put(40,15){\circle*{3}}
           \end{picture}\end{array}
      } 
\bigg] 
~~.
\end{array}
\end{eqnarray}
Define $t=(i+k)/2$ and $\epsilon=(k-i)/2$. The matrix element
$\tilde{W}^{(4)}_{[012]}(a,b,c,d){}_{i}^{k}(a,b,c,d)$ 
is different from zero only if $\epsilon=\pm 1$ and 
all the 3-vertices in equation (\ref{eqWred}) are admissible.
Indeed, by performing the chromatic evaluations of eq.\
(\ref{eqWred}), one gets the 
following explicit formula for the matrix elements
($\epsilon=\pm 1$):
\begin{equation}
\begin{array}{rcll}
\displaystyle
\tilde{W}^{(4)}_{[012]}(a,b,c,d){}_{t-\epsilon}^{t+\epsilon} 
  &=& -\epsilon (-1)^{\frac{a+b+c+d}{2}} 
\\[4mm] &&~~~ 
\displaystyle \bigg[
     \frac{1}{4 t (t+2)} \frac{a+b+t+3}{2}\frac{c+d+t+3}{2}
\\[4mm] &&~~~ \displaystyle
     \frac{1+a+b-t}{2}\frac{1+a+t-b}{2}\frac{1+b+t-a}{2}
\\[4mm] &&~~~ \displaystyle
     \frac{1+c+d-t}{2}\frac{1+c+t-d}{2}\frac{1+d+t-c}{2}
     \bigg]^{\frac{1}{2}}  
\end{array}
\!\!\!\!\!
\end{equation}
This formula can be used to obtain explicit formulas 
for the eigenvalues of the volume operator.
For example, in the case $d=a+b+c-2$, where the matrices 
$\tilde{W}_{[rst]}$ are two-dimensional, one obtains the following
result for the (degenerate) eigenvalue of the volume associated to the
4-vertex $\left\langle V^{(4)}[a,b,c,a+b+c-2] \right|$:
\begin{equation}
v(a,b,c,d) = \frac{l_0^3}{\sqrt{2}} 
             \left[{\frac{a~b~c~(a+b+c)}{16}}\right]^{\frac{1}{4}} ~~,
\end{equation} 
that can be compared with the results of \cite{VOLUME}.

\section{The Normalized State and the Scalar Product}

A scalar product between two spin-network states can be defined by assuming
that it is different
from zero only if the two states have exactly the same
graph and the same coloring of the real edges.
Then, the normalization (\ref{eq:newbasis}), with respect to which the
volume operator is represented by symmetrical matrices, 
determines this scalar product uniquely.  Its value is
determined by the chromatic evaluation of the vertices:
\begin{equation}
  \langle{s},{s'}\rangle 
= \delta_{\gamma,\gamma'} 
 \prod_{e\in {\cal E}_s} \frac{\delta_{n_e,n_e'}}{\Delta_{n_e}} 
 \prod_{i\in {\cal V}_s} \langle V_i , V'_i \rangle  ~~,
\label{eq:normSN} 
\end{equation} 
where the products are extended only to the real edge and to the
real vertices, and $\langle V_i , V'_i \rangle$ is the chromatic
evaluation obtained by gluing the vertices $V_i$ and $V_i'$.  In
\cite{DePietri96a}, it has been shown that the scalar product defined
in this way is precisely the loop transform \cite{FunctINT} of the
Ashtekar-Lewandowski measure \cite{CONNECTION}.

\vspace{1cm}

I thank warmly Massimo Pauri for a critical reading of the manuscript.

%
%

\end{document}